%% file: ms.tex
\documentclass[12pt,preprint]{aastex}
\usepackage{rotating}
\usepackage{float}
\usepackage{longtable}
\usepackage{lscape}
\usepackage{threeparttable}

\begin{document}

\title{Spectroscopic Study of the HST/ACS PEARS Emission-Line Galaxies}

\author{Lifang Xia\altaffilmark{1}, Sangeeta Malhotra\altaffilmark{1},
James Rhoads\altaffilmark{1}, Norbert Pirzkal\altaffilmark{2}, Zhenya Zheng\altaffilmark{1,3}, 
Gerhardt Meurer\altaffilmark{4}, Amber Straughn\altaffilmark{5}, Norman Grogin\altaffilmark{6}, 
David Floyd\altaffilmark{7}}

\altaffiltext{1}{School of Earth and Space Exploration, Arizona State University, AZ, 85287; lifang.xia@asu.edu}
\altaffiltext{2}{Space Telescope Science Institute, Baltimore, MD, 21218; npirzkal@stsci.edu}
\altaffiltext{3}{Center for Astrophysics, University of Science and Technology of China, Hefei, Anhui, 230026, China; zhengzy@mail.ustc.edu.cn}
\altaffiltext{4}{Department of Physics and Astronomy, Johns Hopkins University, Baltimore, MD, 21218; meurer@pha.jhu.edu}
\altaffiltext{5}{NASA Goddard Space Flight Center, Greenbelt, MD, 20771; amber.straughn@asu.edu}
\altaffiltext{6}{Space Telescope Science Institute, Baltimore, MD, 21218; nagrogin@stsci.edu}
\altaffiltext{7}{AAO/OCIW Magellan fellow, School of Physics, University of Melbourne, VIC, 3010, Australia; dfloyd@unimel.edu.au}

\begin{abstract}

  We present spectroscopy of 76 emission-line galaxies (ELGs) in CDF-S taken
  with the LDSS3 spectrograph on Magellan Telescope. These galaxies
  are selected to have emission lines with ACS grism data in the {\it Hubble Space Telescope}
  Probing Evolution and Reionization Spectroscopically (PEARS) grism Survey. 
  The ACS grism spectra cover the wavelength range
  6000-9700 \AA\ and most PEARS grism redshifts are based on a single emission line
  + photometric redshifts from broad-band colors; the Magellan spectra
  cover a wavelength range from 4000 {\AA} to 9000 {\AA}, and provide a
  check on redshifts derived from PEARS data. We find an accuracy of
  $\sigma_z$ = 0.006 for the ACS grism redshifts with only one catastrophic
  outlier. We probe for AGN in our sample via several different methods. In total
  we find 7 AGNs and AGN candidates out of 76 galaxies. Two AGNs are identified from the X-ray full-band
  luminosity, $L_{X-ray,FB}>10^{43}$ erg$\;$s$^{-1}$, the line widths and the power-law continuum spectra.
  Two unobscured faint AGN candidates are identified from the X-ray full-band
  luminosity $L_{X-ray,FB}\sim10^{41}$ erg$\;$s$^{-1}$, the hardness ratio and the column density, 
  and the emission-line and X-ray derived SFRs. Two candidates are classified 
  based on the line ratio of [NII]$\lambda$6584/H$\alpha$ versus [OIII]$\lambda$5007/H$\beta$ 
  (BPT diagram), which are between the empirical and theoretical demarcation curves, i.e, 
  the transition region from star-forming galaxies to AGNs.
  One AGN candidate is identified from the high-ionization emission line HeII{\AA}4686.
  
\end{abstract}


\section{Introduction}

The HST/ACS/G800L grism survey Probing Evolution and Reionization Spectroscopically
(PEARS, PI: S. Malhotra) produces low-resolution (R $\sim$ 100) 
slitless spectra in the wavelength range from 6000{\AA} to 9700{\AA}. 
The survey covers four ACS pointings in GOODS
North (GOODS-N) and five ACS pointings Chandra Deep Field South (CDF-S) fields yielding
spectra of all objects up to $z=27$ magnitude up to $z=28$ magnitude 
in the Hubble Ultra Deep Field (HUDF). We
selected emission-line galaxies in CDF-S from the samples of Xu et
al. (2007), and Straughn et al. (2008, 2009), regardless of the
broad-band magnitude for followup with Magellan telescope for R $\sim$ 1900
spectroscopy. Thus we are able to get spectra for much
fainter objects than have been selected traditionally (e.g. Vanzella
et al. 2006, 2008). One of the aims of the followup spectroscopy is to
confirm the redshifts obtained from the grism data.

The grism data, due to the limited wavelength coverage and low
spectral resolution, often yields only a single unresolved line. For 
single-line spectra, the lines are identified as:
[OII]$\lambda$3727{\AA}, [OIII]$\lambda\lambda$4959,5007{\AA} 
and $H\alpha$ based on photometric
redshifts derived from the broad-band colors (Xu et al. 2007, Straughn
et al. 2008, 2009). 

In this paper, we present the confirmation of the ACS grism redshifts by
the follow-up Magellan LDSS-3 multislit spectroscopic observation of a
sample of 107 emission-line galaxies (ELGs) pre-selected by \citet{straughn09} in the GOODS-S
field. We also compare the flux calibration in the two observations.
The normal star-forming galaxies and AGNs are classified by the
emission-line ratios of the BPT diagnostics diagram \citep{baldwin81}
and X-ray observations.
The paper is organized as below. We briefly describe the observation
and the data reduction in $\S$ 2. The result of redshift comparison
with grism measurement, flux calibration comparison and AGNs
classification are illustrated in $\S$ 3. Finally, we present the
summary in $\S$ 4.

\section{Data and Reduction}

From the HST/ACS PEARS grism survey, \citet{straughn09} selected 203
emission-line galaxies by a 2-dimensional detection and extraction
procedure in the GOODS-S field. The line luminosities of grism
observations extend the studies of star-forming galaxies to
$M\sim-18.5$ at $z\sim1.5$. Starting from 107 pre-selected emission-line 
galaxies, we obtain 89 emission-line galaxies
spectra from the follow-up Magellan LDSS-3 multislit spectroscopic
observation after excluding the undetected spectra and bad spectra.  
With 13 galaxies observed twice, the final sample includes 76 different galaxies.
Figure 1 shows the apparent magnitude distribution of
the total pre-selected ELGs put on masks (dashed line)
and the sample of 76 different galaxies with follow-up spectroscopic observation (solid
line). The pre-selected emission-line galaxies cover magnitude range
from 18.0 to 27.0 with a peak at 23.5. The subsample for follow-up
observation follows the same distribution.

\begin{figure}[htbp]
\begin{center}
\figurenum{1} \epsscale{0.7}
\hspace{-1.0cm}
{\plotone{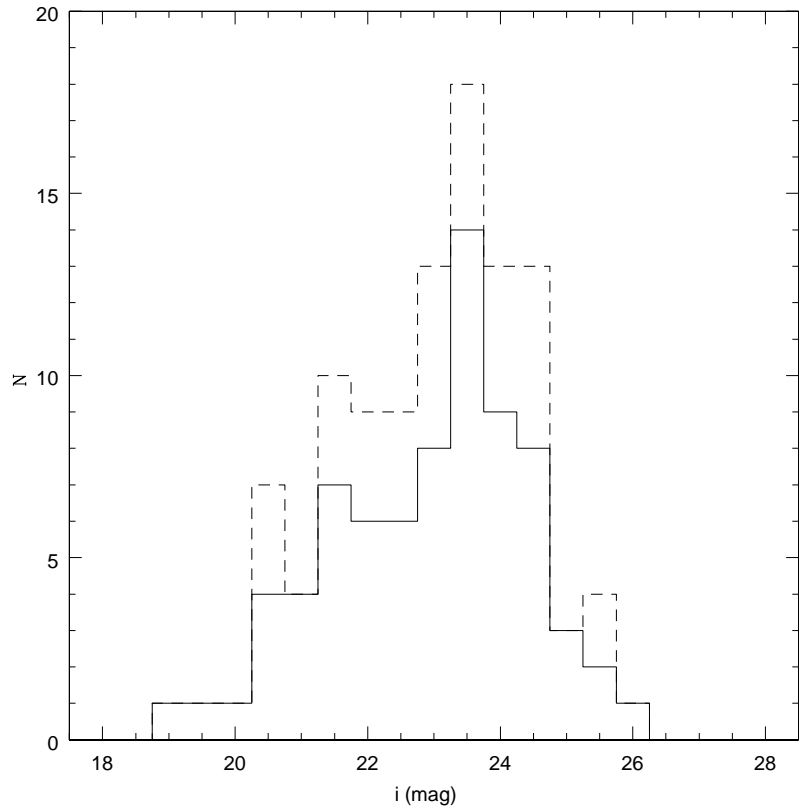}}
\caption{Histogram of the distribution of $i$-band (ACS F775W) apparent magnitudes. 
The dashed line is that of the total 107 emission-line galaxies put on masks. The solid line 
is that of the 76 emission-line galaxies with good quality redshift measurements from LDSS3.
The magnitudes of the sample peak at $i=23.5$.}
\end{center}
\end{figure}

The spectroscopic follow-up was done in a total of four nights in
November 2007 and December 2008 using the Magellan LDSS-3 spectrograph
and using the VPH-Blue and VPH-Red grisms. The LDSS-3 instrument has a
scale of 0$\farcs$189/pixel. The VPH-Blue grism covers the
wavelength range from 4000{\AA} to 6500{\AA} with a resolution of
$R=1810$, dispersion of 0.682{\AA}/pixel@5200{\AA}. The VPH-Red grism
covers the wavelength range from 6000{\AA} to 9000{\AA} with OG590
filter used to eliminate contamination from the second order. The red
grism has a resolution of $R=1900$ and dispersion of
1.175{\AA}/pixel@8500{\AA}. We used slit widths of
0$\farcs$8. 

Five masks were created to contain all of the
science objects with 4-6 alignment stars located at different parts of
each mask. The fields were observed with integration times of 5400s,
7200s, and 8100s.  For masks observed in 2007, the spectroscopic
standard star LTT1020 was observed for calibration; in 2008, the
spectroscopic standard stars, LTT1020, LTT2415, EG21 and LTT3864 were
observed for flux calibration.

We reduced the spectra using the {\it COSMOS} software package 
\citep{cosmos09}, which is designed for multislit spectra obtained using
the IMACS and LDSS3 spectrographs on Magellan.
Following the reduction process of making
alignment, subtracting bias, flattening, wavelength calibration, sky
subtraction and 2-dimensional spectra extraction, the blue-end and
red-end spectra were obtained for all objects. The 1-d spectra
extraction and flux calibration were accomplished in $IRAF$.

To check the flux calibration from year to year we compared the
calibrated spectra for objects observed in both years. Upon doing
this, we realized that the flux calibration of 2007 data, which was
based on a single calibration star was systematically higher. This, we
conjectured, must be due to misplacement of the standard star in the
slit. The sensitivity function of the CCD obtained from the spectroscopic 
standard stars observation in 2008 is applied to the flux calibration 
of the 2007 data. To check its robustness, we then used object 110494, 
which has strong continuum and is
observed in both years. Figure 2 shows the two flux calibrated spectra
for the object. The blue and red spectra are combined together to
cover wavelength range from 4000{\AA} to 9000{\AA}. The spectra show
consistency in the junction point at 6500{\AA} of the blue and red
ends. The dotted line shows the spectra obtained from 2007 data and the
solid line represents that of 2008. The main strong emission lines
emerging in the spectra are [OII]$\lambda$3727, H$\beta$,
[OIII]$\lambda\lambda$4959,5007, and H$\alpha$.  We fit the continuum
of the two spectra and find a difference of 5\% in the continuum flux
from 5000 {\AA} to 9000 {\AA}. We measure the line fluxes and 
errors for H${\beta}$ and H${\gamma}$, and obtain the ratio of 
H${\gamma}$/H${\beta}$ = 0.45$\pm$0.05, and 0.48$\pm$0.07, separately. The ratios 
are in good agreement with each other and with the theoretical 
value, 0.469. The good agreement of the continuum and the line ratios of the two years
spectra demonstrate that the calibration is sufficiently robust for
our purpose.

From the 2-dimensional spectra, we finally obtained 89 sources which
show clear detection of emission lines. The galaxy redshifts are first
visually determined from the pattern of the emission lines. The
accurate redshifts and uncertainties are determined by
the average and variance of the redshifts obtained from the main
emission lines in the spectra. In the 89 spectra, there are 13 objects
which were observed in both years.  We finally obtain 76
unique redshifts which are used to assess the accuracy of the grism
redshifts at $0.1<z<1.3$.

Excluding objects only observed in blue or red end, objects with signal to noise ratio less than 3 in H$\beta$,
and [OIII]$\lambda\lambda$4959,5007, and objects with one or more emission lines out of spectral coverage, 
we measure the line fluxes for 55
well extracted 1-d spectra with whole set of [OII]$\lambda\lambda$3727,3729,
H$\beta$, and [OIII]$\lambda\lambda$4959,5007 lines. The emission-line
fluxes are measured by Gaussian fitting ({\sc gaussfit} in {\sc idl})
expanding 40{\AA} around the line peak. Most of the FWHM of the line profiles are 
in the range from 2 {\AA} to 9 {\AA} with line velocities $<$ 500 km s$^{-1}$, 
except two objects, 92839 and 102156, of 28 and 79 {\AA}, corresponding to velocities
$\sim$ 1000, 3800 km s$^{-1}$ (discussed in $\S$ 3.3).

\begin{figure}[htbp]
\begin{center}
\figurenum{2} \epsscale{1.0}
\hspace{-1.0cm}
{\plotone{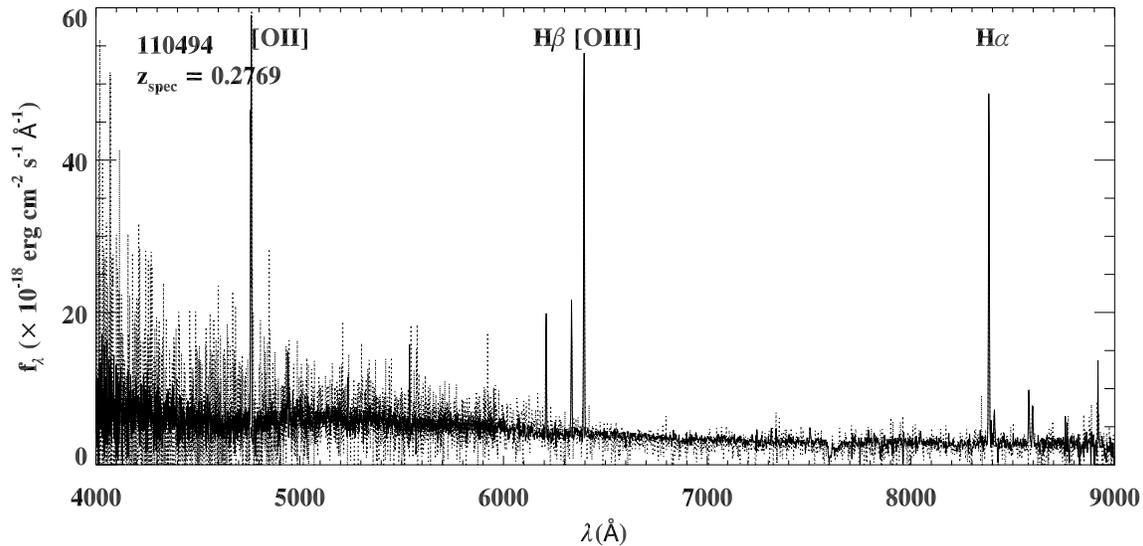}}
\caption{Flux calibrated spectra for object 10494 observed in both
2007 (dotted line) and 2008 (solid line). The flux uncertainties of the spectra
in 2007 are much larger than that of 2008 due to the larger seeing.) 
Due to the off-slit positioning of of the standard star 
in 2007 data, the spectra of 2007 is flux calibrated by the sensitivity function 
obtained from 2008 spectroscopic standard stars. The consistency 
of the continuum and the line ratio of H$\gamma$/H$\beta$ in the two years demonstrate 
the robustness and effectiveness of this application. The PEARS ID, the redshift and the main emission 
lines are labeled in the plot.}
\end{center}
\end{figure}

\section{Results}

Table 1 lists the general information and the measurement results of the galaxy sample, the 
PEARS ID (column 1), R.A. (column 2), Dec. (column 3), 
$i$ magnitude (column 4), spectroscopic redshifts (column 5), grism redshifts (column 6), 
the FWHM of line H$\beta$ (column 7),
the flux and flux error of [OIII]$\lambda$4959,5007 in the Magellan spectroscopy (column 8) 
and the PEARS grism survey (column 9).

\subsection{Redshift Comparison}

We first compare the LDSS3 redshifts with the redshifts
determined from ACS grism detections of 1 or 2 emission lines 
at $\sim$80{\AA} resolution. Among the 76 emission-line galaxies with
LDSS3 redshifts, 62 have ACS grism redshifts from \citet{straughn09}. 
For remaining 14 \citet{straughn09}
find a line but cannot assign a line identification and redshift with
confidence due to lack of secure photometric redshift for these sources.
We plot the redshift differences between the LDSS3 and ACS 
redshifts in Figure 3. The ACS grism redshifts include only one
catastrophic failure (object 89030, discussed below) and one object, 72509, 
with redshift difference of 0.05. Object 72509 has a redshift of 1.246 and only
the [OII]3727 is observed in the red-end of the spectra. The ACS grism
spectrum of this object is noisy and there are several peaks around 8400 {\AA} 
which could be due to the contamination of sky line residuals.
Among the remaining 60 objects, we measure a root mean square redshift difference of
$\sigma_z=0.006$ between the ACS and LDSS3 redshifts. 

Object, 89030, with large deviation between the measured
spectroscopic redshift, 0.6220, and the grism redshift, 1.449, has a well detected
continuum, $f_{\lambda} \sim 10^{-18}$ erg$\;$s$^{-1}\;$cm$^{-2}\;${\AA}$^{-1}$, and a full set of lines,
[OII] doublet, H$\beta$, and [OIII] doublet, in the Magellan spectrum. 
The ACS grism spectrum has the strongest line peaks around 9120{\AA}, 
which is assigned to be [OII]$\lambda$3727, and a weak continuum 
$f_{\lambda} \sim 10^{-19}$ erg$\;$s$^{-1}\;$cm$^{-2}\;${\AA}$^{-1}$. From the $i$-band image 
of this object, it is found that object 89030 has two neighbors, an extended spiral 
and a bright compact object. Combined with the faint $i$-band magnitude, $i=25.79$, 
we conclude that the spectrum obtained from Magellan could be the
contamination of one of the adjacent two objects.

\begin{figure}[htbp]
\begin{center}
\figurenum{3} \epsscale{0.9}
\hspace{-1.0cm}
{\plotone{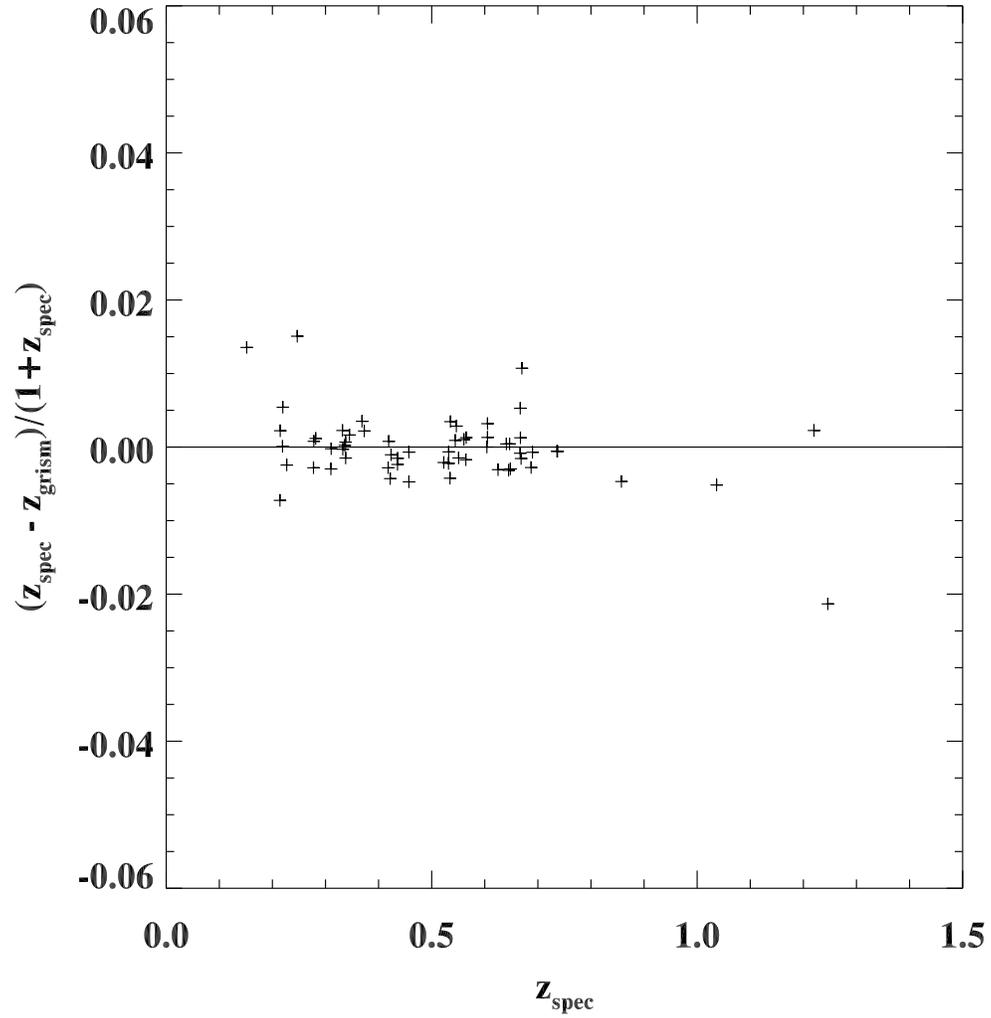}}
\caption{Redshift differences between the spectroscopic and the grism redshifts as a function of the spectroscopic redshifts. 
The accuracy of the grism redshift is measured to be $\sigma_z=0.006$.}
\end{center}
\end{figure}


\subsection{Flux Comparison}

We compare emission-line fluxes as measured from the ground and the
grism. Usually, [OIII]$\lambda$5007 is the strongest emission line in
the spectra. Due to the low resolution of ACS grism spectra, the two lines
[OIII]$\lambda\lambda$4959,5007 are blended into one wide peak. Figure 4
presents the comparison of the total emission-line fluxes of
[OIII]$\lambda\lambda$4959,5007 for 33 common objects with both flux
measurements. The $y$-axis is the flux ratio between the spectroscopic
to the grism flux and the $x$-axis is the geometric mean of the grism and the
spectroscopic line fluxes. From the figure, the ratio for most of the galaxies are in the
range from 0.5 to 2 (dotted line), which agrees with the
expectation. In the pre-selected ELGs sample about two-thirds have
irregular and/or merging morphologies \citep{straughn09}. For 
irregular and extended morphologies the slit losses can lead to a factor of 2
underestimatation of the spectroscopic line fluxes. The ACS grism spectra
are extracted for individual star forming knots based on the 2D detection \citep{straughn09},
which could introduce big differences for flux comparison also. Other factors,
such as the uncertainty in the background continuum determination of the ACS spectra,  
the contamination of the H$\beta$ can introduce some factor to the line fluxes. Therefore, we assume
that the factor from 0.5 to 2 in the flux ratio is in the reasonable
range of the measurements.

\begin{figure}[htbp]
\begin{center}
\figurenum{4} \epsscale{1.0}
\hspace{-1.0cm}
{\plotone{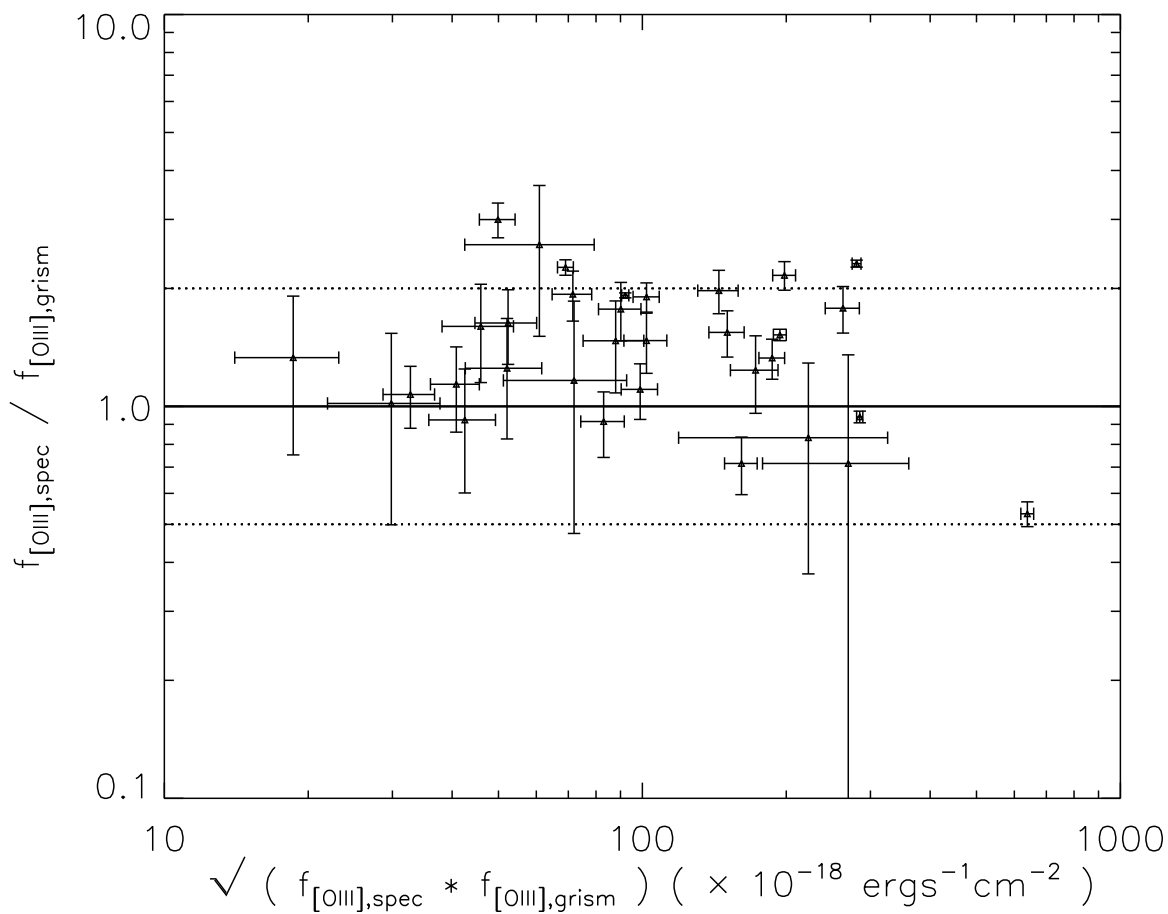}}
\caption{Flux ratios of the spectroscopic to the grism as a function of the square root of 
the [OIII] line fluxes measured by ACS grism and LDSS-3, which is plotted in log scale. The ratios for most objects are in the range
from 0.5 to 2.0 (the dotted lines, the solid line shows the ratio of 1), 
which is in the reasonable range due to the different sampling of galaxy 
light by the slit and grism, the uncertainty in the determination of the grism continuum.}
\end{center}
\end{figure}

\subsection{AGN Identification}

The contribution to the emission lines in spectra includes the ionized
HII region by massive stars in normal star-forming galaxies and the
narrow-line region (NLR) of AGNs. To classify the emission-line
galaxies in our sample to be star-forming galaxies or AGNs, we use 
two methods: catalog matching to the CDF-S X-ray sources catalog of \citet{luo08},
and comparison of the [NII]$\lambda$6584/H$\alpha$ versus [OIII]/H$\beta$
line ratios (i.e. the well known BPT diagram; \citep{baldwin81}. The cross-check
with the X-ray detections gives 5 X-ray counterparts with separation
within 2$^{\prime\prime}$, which are possible AGNs and are marked in
Table 1. By checking the X-ray full-band flux, the two objects, 92839
and 102156, have a luminosity of $L_{FB}=6.36\times$10$^{43}$ ergs
s$^{-1}$ and 3.36$\times$10$^{43}$ erg$\;$s$^{-1}$, respectively. From
the spectra, these two objects show strong exponential-slope
continuum. From the line widths, the lines of these two AGNs 
have velocities $\sim$ 3800 km$\;$s$^{-1}$, $\sim$ 1000 km$\;$s$^{-1}$.
Thus, these two are determined to be broad-line AGNs. 

The other three objects, 59018, 60143, and 79483, show $L_{FB}\sim10^{41}$ ergs$\;$s$^{-1}$ 
and are possible starburst galaxies and faint AGNs.
We derive the hardness ratios, HR=(H-S)/(H+S), where S and H are 
counts in the soft-band (0.5-2 keV) and in the hard-band (2-7 keV), 
for the two galaxies, 59018 and 79483. The HRs are $<$ -- 0.13 and 
$<$ -- 0.29, respectively, which implies an intrinsic absorption 
of X-ray column density N$_H <$ 8.8 and 2.4 $\times 10^{21}$ cm$^{-2}$ 
(68\% confidence level, for $\gamma = 2.0$ and solar metallicity). 
This suggests that the X-ray fluxes are dominated by star formation 
or unobscured faint AGN. 

We use the extinction corrected (the extinction is obtained by the continuum SED fitting with 
the BC03 stellar population synthesis model, Bruzual \& Charlot 2003) line fluxes
of [OII] and $H\beta$ to derive the star formation rates (SFR) for the 
three possible starburst galaxies by the
calibrations given by \citet{kennicutt98}, and use the soft-band
(0.5-2 kev) and hard-band (2-10 kev) X-ray fluxes to get SFR 
by the relations given by \citet{ranalli03}. The results are given
in Table 2. The ``$<$'' in Table 2 denotes the upper limit X-ray 
detection. The X-ray flux of galaxy 60143 is only 
detected in the full band (0.5-7 keV). 
The SFRs of object 60143 agree very well between the 
[OII]-derived and soft-band derived results, $\sim$ 10 M$_\sun$/yr, so galaxy 
60143 are more likely a starburst galaxy. 
For object 59018 and 79483, the X-ray 
calibrations give the SFR $\sim$ 10 M$_\sun$/yr, and the emission lines
calibrations give the SFR $\sim$ 1 M$_\sun$/yr. While the SFRs from 
X-ray are an order larger than the SFRs from the extinction-corrected 
emissions for galaxies 59018 and 79483, we treat these two galaxies 
as unobscured faint AGNs.

For the emission-line sources, the lines H$\alpha$ and
[NII]$\lambda$6584 can only be observed for galaxies at $z<0.36$ due
to the wavelength coverage of our spectra. The above 5 objects with
X-ray detection all have redshift $z>0.36$ and hence out of the
analysis of the BPT diagonostic method. For 14 galaxies with good line
flux measurements at $z<0.36$, Figure 5 shows the plot of the
[NII]$\lambda$6584/H$\alpha$ and [OIII]/H$\beta$ ratios for these
objects. The theoretical maximum starburst limit (dashed line) from
\citet{kewley01} and the empirical demarcation from
\citet{kauffmann03} (dotted line) are also plotted. All of the 14
objects are below the theoretical upper limit \citep{kauffmann03}.  
Two object, 89923 and 111549,
lie in the transition region between the empirical and theoretical
demarcation curves. There are no X-ray detections for these two objects, no
other distinct AGN high ionization indictator emission lines,
e.g. [NeV] and HeII, and no broad lines. Hence, these objects could be 
star-forming galaxies, or low-luminosity AGNs, or some combination
of the two. 

For galaxies at $z>0.36$ and without H$\alpha$ and [NII] observation,
we use the HeII$\lambda$4686 as the indicator of the AGN activity.
Only one object, 106761, has prominent HeII in the spectra and could be
AGN. 

The above analysis of the X-ray detection, line width, hardness ratio
and column density, SFRs, BPT diagram and high ionization emission line,
give 7 AGNs in our sample. We mark these objects in Table 1 with stars 
besides the object ID as the AGNs and AGN candidates identified in this paper.

\begin{figure}[htbp]
\begin{center}
\figurenum{5} \epsscale{1.0}
\hspace{-1.0cm}
{\plotone{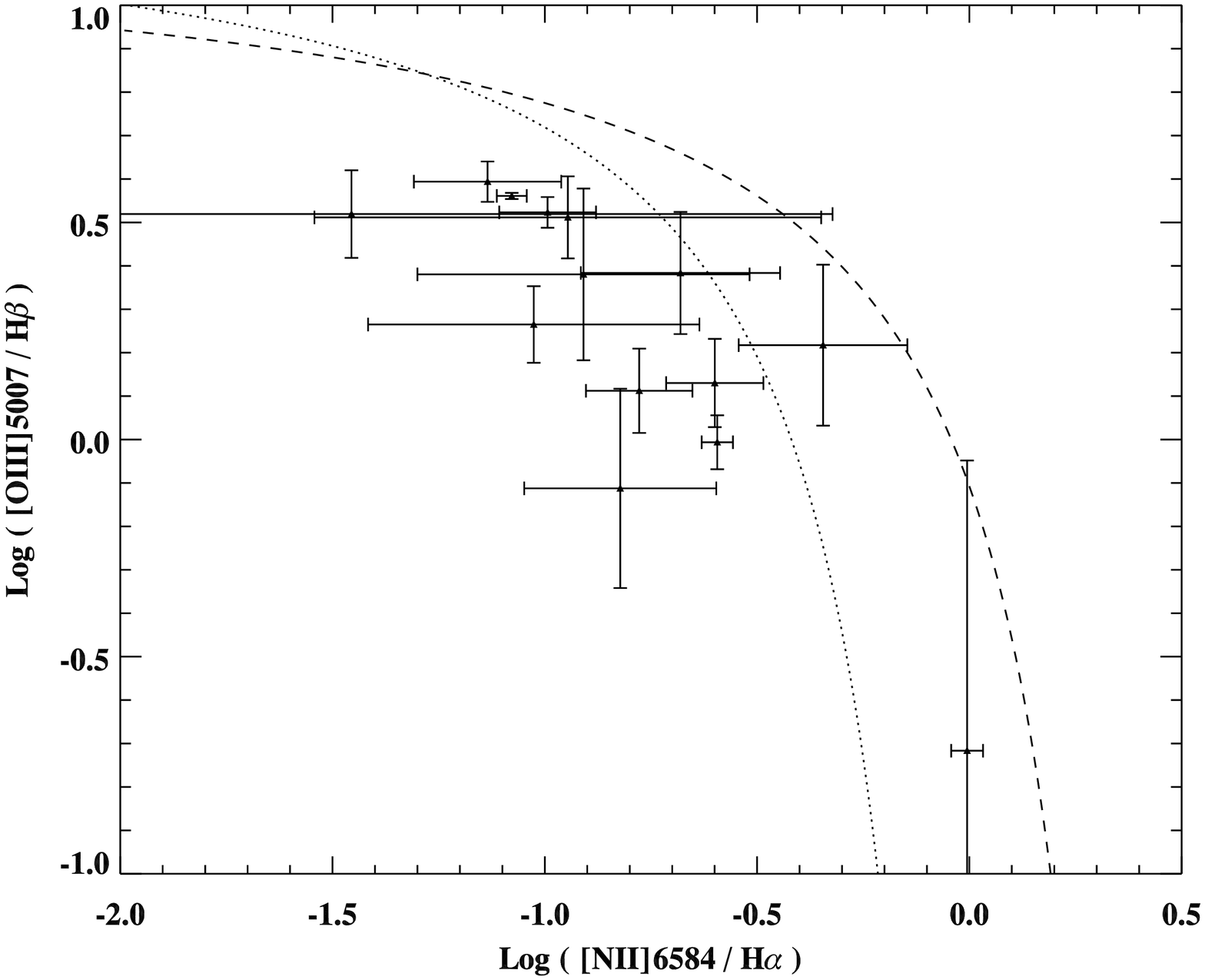}}
\caption{Emission-line ratios [NII]$\lambda$6584/H$\alpha$ vs. [OIII]/H$\beta$ for 14 objects at $z<0.36$ with H$\alpha$ and [NII]$\lambda$6584
observation and measurements. The dashed line is the theoretical maximum starburst limit from \citet{kewley01}, and the dotted line represents the
empirical demarcation from \citet{kauffmann03} (dotted line). Two objects in the locus between the two curves have large [NII]$\lambda$6584 
line flux and have high probability to be AGNs.}
\end{center}
\end{figure}

\section{Summary}

We investigate the accuracy of the grism redshifts using the Magellan
LDSS-3 follow-up spectroscopic observation of a sample of 76 emission-line 
galaxies. The galaxies are pre-selected to have emission lines
\citep{straughn09} in the GOODS-S field. The galaxies span the
magnitude range $19.0<i<26.0$ and the redshift range $0.1<z<1.3$. In
the spectral coverage from 6500{\AA} to 9700{\AA}, the most important
emission line observed are [OII], H$\beta$, [OIII], and some
H$\alpha$, and [NII] for low redshift galaxies. The spectroscopic
redshifts are measured from the pattern of the emission lines. The
spectroscopic redshifts of 76 galaxies are obtained. The accuracy of the
grism redshifts is assessed using 62 galaxies with both redshift
measurements. An accuracy of $\sigma_z=0.006$ is found for the grism redshifts.

For 33 galaxies with both LDSS-3 flux measurements and grism fluxes,
the emission-line fluxes of [OIII] are compared. A general agreement
is found with the [OIII] flux ratio ranging from 0.5 to 2. The different 
sampling of light by the slit and the ACS grism, and the
uncertainty in the continuum determination of the ACS grism spectra may
result in this factor of 2.

By cross-checking with CDF-S X-ray catalog \citep{luo08}, two AGNs, 92839 and 102156, are
identified with luminosities of $L_{FB}>10^{43}$ erg$\;$s$^{-1}$. 
Another three X-ray detected galaxies show luminosity of
$L_{FB}\sim10^{41}$ ergs s$^{-1}$ and are possible starburst galaxies or obscured faint AGNs.
The SFRs for the three objects are derived from extinction corrected emission-line 
fluxes and X-ray soft-band and hard-band fluxes. One object, 60143,
shows good agreement in the derived-SFRs, which is $\sim$ 10 M$_\sun$/yr,
and is more likely a starburst galaxy. 
For another two galaxies, 59018 and 79483, the hardness ratio, HR $<$ -- 0.13 and $<$ -- 0.29,
and the X-ray column density, N$_H <$ 8.8 and 2.4 $\times 10^{21}\;$cm$^{-2}$, 
suggests possible star formation or unobscured faint AGNs. Since
the extinction corrected emission-line [OII] and H$\beta$ derived SFRs are $\sim$ 1 M$_\sun$/yr,
while the X-ray derived SFR is $\sim$ 10 M$_\sun$/yr, we treat these two
galaxies as unobscured faint AGNs.

For 14 galaxies at $z<0.36$ (without X-ray counterparts) and with H$\alpha$ and [NII]
emission lines observed in the spectra, we use the BPT
diagram to identify star-forming galaxies and AGNs. All of the 14 objects
locate below the theoretical upper limit \citep{kauffmann03}.
Two objects, 89923 and 111549, locating in the transition region between
star-forming galaxies and AGNs, could be possible AGNs.
From the high ionization indictator emission lines, HeII$\lambda$4686, one more object, 106761,
is identified as possible AGN.

\begin{acknowledgements}

This paper includes data gathered with the 6.5 meter Magellan Telescopes located at Las Campanas Observatory, Chile.
PEARS is an HST Treasury Program 10530 (PI: Malhotra). Support for program was provided
by NASA through a grant from the Space Telescope Science Institute, which is operated by
the Association of Universities for Research in Astronomy, Inc., under NASA contract
NASA5-26555 and is supported by HST grant 10530.

\end{acknowledgements}

\clearpage

\input{t1.tex}

\input{t2.tex}

\end{document}

%% file: t1.tex



\scriptsize
\begin{longtable}{lcccccccc}
\caption[Redshifts and fluxes]{Spectroscopic redshifts and emission line fluxes of the emission line galaxies obtained from the Magellan
follow-up LDSS-3 observation. The corresponding grism redshifts and grism fluxes are listed in the table. The stars besides object ID 
represent AGNs and AGN candidates.} \label{z_flux} \\[0.5ex]

\hline\hline \\[-2ex]
\multicolumn{1}{c}{PEARS ID} & \multicolumn{1}{c}{RA} & \multicolumn{1}{c}{DEC} & \multicolumn{1}{c}{$i_{mag}$\tablenotemark{a}} & 
\multicolumn{1}{c}{$z_{spec}$} & \multicolumn{1}{c}{$z_{grism}$}  & \multicolumn{1}{c}{$FWHM$\tablenotemark{b}} & \multicolumn{1}{c}{$f_{[OIII],spec}$\tablenotemark{c}} & 
\multicolumn{1}{c}{$f_{[OIII],grism}$\tablenotemark{c}}
\\[0.5ex] \hline
\\[-1.8ex]
\endfirsthead

\multicolumn{8}{l}{{\tablename} \thetable{} -- Continued} \\[0.5ex]
\hline \hline \\[-2ex]
\multicolumn{1}{c}{PEARS ID} & \multicolumn{1}{c}{R.A.} & \multicolumn{1}{c}{DEC} & \multicolumn{1}{c}{$i_{mag}$\tablenotemark{a}} & 
\multicolumn{1}{c}{$z_{spec}$} & \multicolumn{1}{c}{$z_{grism}$} & \multicolumn{1}{c}{$FWHM$\tablenotemark{b}} & \multicolumn{1}{c}{$f_{[OIII],spec}$\tablenotemark{c}} & 
\multicolumn{1}{c}{$f_{[OIII],grism}$\tablenotemark{c}}
\\[0.5ex] \hline
\\[-1.8ex]
\endhead

\\[-1.8ex] \hline \hline
\multicolumn{8}{l}{{Continued on Next Page\ldots}} \\
\endfoot

\\[-1.8ex] \hline \hline
\endlastfoot

     12250   &   3:32:37.61    &  -27:55:32.63     &  24.69   &   0.3391   &     --       &   6.1  &    100.5$\pm$10.7     &          --          \\
     13541   &   3:32:38.03    &  -27:55:08.07     &  21.41   &   0.3730   &    0.370     &   4.8  &    191.9$\pm$10.6     &     155.1$\pm$34.3   \\
     17587   &   3:32:38.60    &  -27:54:49.85     &  24.81   &   0.6447   &    0.650     &   1.2  &     58.3$\pm$8.5      &      46.6$\pm$15.5   \\
     17686   &   3:32:27.87    &  -27:54:51.56     &  29.73   &   0.6697   &     --       &    --  &         --            &          --          \\
     18862   &   3:32:32.72    &  -27:54:22.91     &  19.24   &   0.2018   &     --       &   3.3  &     82.6$\pm$35.8     &          --          \\
     19422   &   3:32:41.30    &  -27:54:34.74     &  24.51   &   0.5506   &    0.553     &   4.9  &    104.0$\pm$9.9      &      94.1$\pm$13.7   \\
     19639   &   3:32:34.92    &  -27:54:13.83     &  19.90   &   0.2802   &    0.280     &   3.0  &    178.9$\pm$17.2     &     125.0$\pm$83.3   \\
     22829   &   3:32:39.54    &  -27:54:00.67     &  21.52   &   0.5606   &    0.559     &   5.0  &    239.3$\pm$13.4     &     157.2$\pm$4.3   \\
     26009   &   3:32:33.10    &  -27:53:40.68     &  23.60   &   0.4356   &    0.439     &    --  &         --            &          --          \\
     31362   &   3:32:43.68    &  -27:53:05.90     &  24.17   &   0.6672   &    0.665     &   4.8  &    275.9$\pm$0.5      &     293.8$\pm$8.7   \\
     33294   &   3:32:38.08    &  -27:52:48.68     &  23.49   &   1.0354   &    1.047     &    --  &         --            &          --          \\
     37690   &   3:32:40.74    &  -27:52:16.92     &  23.57   &   0.3644   &     --       &   1.8  &     44.9$\pm$3.7      &          --          \\
     41078   &   3:32:43.39    &  -27:51:54.54     &  24.25   &   0.8573   &    0.866     &    --  &         --            &          --          \\
     43170   &   3:32:37.49    &  -27:51:38.84     &  24.02   &   0.6874   &    0.692     &   7.0  &    123.4$\pm$4.3      &      84.1$\pm$15.6   \\
     45454   &   3:32:43.63    &  -27:51:22.37     &  22.73   &   0.4233   &    0.425     &   4.1  &     43.6$\pm$0.1      &      38.2$\pm$9.0   \\
     46994   &   3:32:39.45    &  -27:51:13.16     &  24.29   &   0.6665   &    0.668     &   6.7  &    187.1$\pm$7.1      &     121.1$\pm$17.6   \\
     49766   &   3:32:42.00    &  -27:50:51.80     &  23.53   &   0.2184   &    0.213     &   1.9  &     30.1$\pm$10.9     &          --          \\
     52086   &   3:32:37.87    &  -27:50:39.52     &  23.47   &   0.5227   &    0.526     &   4.9  &    145.0$\pm$8.6      &     243.9$\pm$46.5   \\
     54022   &   3:32:41.93    &  -27:50:26.81     &  22.29   &   0.3360   &    0.336     &   5.6  &    120.0$\pm$3.0      &      67.7$\pm$13.7   \\
     55102   &   3:32:42.15    &  -27:50:18.71     &  21.83   &   0.4567   &    0.458     &   3.8  &     77.7$\pm$6.1      &      66.7$\pm$38.1   \\
     56801   &   3:32:34.82    &  -27:50:14.56     &  23.93   &   0.6491   &    0.653     &   4.3  &     45.6$\pm$7.1      &          --          \\
     56875   &   3:32:36.72    &  -27:50:15.70     &  24.48   &   0.5346   &    0.541     &   4.1  &     34.6$\pm$2.7      &      32.6$\pm$3.3   \\
     58985   &   3:32:47.98    &  -27:50:02.64     &  23.78   &   0.5650   &    0.563     &    --  &         --            &          --          \\
     59018 $^{\star}$\tablenotemark{d}   &   3:32:42.32    &  -27:49:50.33     &  20.59   &   0.4571   &    0.464     &  5.8 &     20.2$\pm$3.4      &          --          \\
     60143   &   3:32:35.61    &  -27:49:43.95     &  21.21   &   0.5464   &    0.542     &   -- &         --            &          --          \\
     65825   &   3:32:41.22    &  -27:49:18.45     &  23.51   &   0.9329   &     --       &    --  &         --            &          --          \\
     70651   &   3:32:36.75    &  -27:48:43.51     &  23.33   &   0.2143   &    0.212     &   3.1  &   179.1$\pm$27.5      &     102.9$\pm$15.4   \\
     72509   &   3:32:40.92    &  -27:48:23.73     &  24.46   &   1.2461   &    1.294     &    --  &        --            &          --          \\
     72557   &   3:32:32.19    &  -27:48:24.41     &  23.52   &   0.3378   &     --       &    --  &        --            &          --          \\
     73619   &   3:32:44.26    &  -27:48:18.58     &  24.77   &   0.6699   &    0.652     &    --  &        --            &          --          \\
     75506   &   3:32:35.34    &  -27:48:03.06     &  26.33   &   0.2794   &    0.277     &    --  &    33.9$\pm$6.9      &      31.6$\pm$4.4   \\
     75753   &   3:32:44.97    &  -27:47:39.22     &  21.57   &   0.3451   &    0.343     &   4.9  &   291.4$\pm$1.0      &     134.9$\pm$14.8   \\
     76154   &   3:32:36.29    &  -27:47:55.32     &  23.68   &   0.6049   &    0.600     &   5.2  &    34.1$\pm$11.3      &      66.4$\pm$0.7   \\
     79283   &   3:32:34.11    &  -27:47:12.10     &  20.75   &   0.2266   &    0.230     &   4.1  &    34.3$\pm$3.9      &          --          \\
     79483 $^{\star}$\tablenotemark{d}   &   3:32:45.11    &  -27:47:24.00     &  20.81   &   0.4345   &    0.438     &  5.9 &     13.5$\pm$1.9      &          --          \\
     80500   &   3:32:35.32    &  -27:47:18.53     &  23.34   &   0.6677   &    0.658     &   4.5  &     66.9$\pm$12.0     &      41.0$\pm$9.6   \\
     81944   &   3:32:34.73    &  -27:47:07.62     &  22.48   &   0.2469   &    0.228     &   3.5  &    525.9$\pm$9.7      &     875.7$\pm$37.8   \\
     83381   &   3:32:42.37    &  -27:46:57.17     &  24.92   &   0.3318   &    0.329     &    --  &         --            &          --          \\
     85517   &   3:32:42.32    &  -27:46:51.06     &  24.79   &   0.5358   &    0.530     &   7.2  &     65.5$\pm$5.0      &          --          \\
     89030   &   3:32:38.50    &  -27:46:30.82     &  25.79   &   0.6220   &    1.449     &   5.0  &     15.8$\pm$6.8      &          --          \\
     89853   &   3:32:33.02    &  -27:46:08.76     &  21.63   &   0.3689   &    0.364     &    --  &         --            &          --          \\
     89923 $^{\star}$\tablenotemark{d}   &   3:32:41.76    &  -27:46:19.39     &  21.25   &   0.3331   &    0.333     &   5.4 &      9.7$\pm$6.4   &          --          \\
     90116   &   3:32:46.76    &  -27:46:24.05     &  25.45   &   0.6250   &    0.630     &    --  &         --            &          --          \\
     91205   &   3:32:36.13    &  -27:46:16.37     &  23.18   &   0.2178   &     --       &   4.2  &    87.3$\pm$18.4      &          --          \\
     91789   &   3:32:35.29    &  -27:46:12.21     &  23.80   &   0.5313   &    0.533     &   4.2  &     21.0$\pm$5.2      &          --          \\
     92839 $^{\star\star}$\tablenotemark{e}   &   3:32:39.08    &  -27:46:01.78     &  20.95   &   1.2222   &    1.215     & 79.\tablenotemark{f} &        --            &          --          \\
     95471   &   3:32:42.56    &  -27:45:50.16     &  22.38   &   0.2191   &    0.219     &    --  &         --            &          --          \\
     96123   &   3:32:34.30    &  -27:45:49.21     &  23.12   &   0.5313   &    0.535     &   4.1  &     21.0$\pm$5.3      &          --          \\
     96627   &   3:32:40.91    &  -27:45:40.91     &  21.50   &   0.1516   &    0.136     &   4.1  &   288.0$\pm$40.1      &          --          \\
     97655   &   3:32:27.37    &  -27:45:40.61     &  23.71   &   0.5442   &    0.543     &   5.0  &    37.5$\pm$10.4      &     589.2$\pm$23.2   \\
    100188   &   3:32:24.31    &  -27:45:24.41     &  25.00   &   0.3107   &    0.311     &    --  &         --            &          --          \\
    102156 $^{\star\star}$\tablenotemark{e}   &   3:32:30.22    &  -27:45:04.60     &  21.65   &   0.7368   &    0.738     & 28. &   228.0$\pm$9.8      &     318.8$\pm$15.2   \\
    104408   &   3:32:27.85    &  -27:44:49.96     &  24.27   &   0.7371   &    0.737     &   5.0  &     97.9$\pm$8.6      &      37.9$\pm$22.6   \\
    105723   &   3:32:27.30    &  -27:44:28.68     &  20.03   &   0.2142   &    0.223     &    --  &         --            &          --          \\
    106491   &   3:32:27.28    &  -27:44:37.46     &  24.93   &   0.3372   &    0.337     &   5.7  &   110.4$\pm$13.5      &      72.5$\pm$20.7   \\
    106761 $^{\star}$\tablenotemark{d}   &   3:32:29.12    &  -27:44:38.63     &  25.88   &   0.6673   &     --       &     2.2 &     54.9$\pm$12.7      &      52.4$\pm$3.2   \\
    109547   &   3:32:21.41    &  -27:44:09.59     &  23.64   &   0.3627   &    0.368     &    --  &         --            &          --          \\
    110494   &   3:32:25.91    &  -27:44:01.49     &  21.96   &   0.2775   &    0.281     &   3.8  &   332.8$\pm$18.1      &     197.0$\pm$31.9   \\
    111549 $^{\star}$\tablenotemark{d}   &   3:32:24.60    &  -27:43:46.79     &  22.06   &   0.3096   &    0.314     &   4.6 &    58.1$\pm$7.8      &      36.3$\pm$11.3   \\
    114392   &   3:32:22.95    &  -27:43:33.09     &  23.63   &   0.5636   &    0.567     &   2.9   &    30.0$\pm$8.7      &      29.6$\pm$13.0   \\
    117138   &   3:32:17.36    &  -27:43:07.27     &  21.18   &   0.6480   &     --       &   2.9   &    92.4$\pm$6.1      &      51.5$\pm$9.2   \\
    117686   &   3:32:18.25    &  -27:43:10.95     &  24.44   &   0.6693   &     --       &   4.5   &    64.8$\pm$11.0      &      28.1$\pm$4.1   \\
    117929   &   3:32:29.52    &  -27:43:05.19     &  22.09   &   0.3378   &    0.340     &   3.2   &    86.4$\pm$6.4      &      28.8$\pm$4.4   \\
    118014   &   3:32:23.68    &  -27:43:08.72     &  23.60   &   0.9796   &     --       &    --   &        --            &          --          \\
    118100   &   3:32:16.87    &  -27:43:04.27     &  23.16   &   0.6467   &    0.646     &   7.2   &   140.8$\pm$10.6      &      74.0$\pm$7.4   \\
    118673   &   3:32:21.94    &  -27:43:03.41     &  24.62   &   0.7362   &     --       &    --   &        --            &          --          \\
    119341   &   3:32:16.81    &  -27:42:59.76     &  25.09   &   0.6909   &    0.691     &   6.2   &    56.3$\pm$10.9      &          --          \\
    121817   &   3:32:23.16    &  -27:42:39.98     &  24.48   &   0.6683   &    0.671     &   4.0   &    79.4$\pm$11.3      &      86.7$\pm$12.9   \\
    123008   &   3:32:16.65    &  -27:42:32.71     &  23.21   &   0.6410   &    0.640     &   5.0   &   215.4$\pm$5.9      &     162.1$\pm$19.4   \\
    123301   &   3:32:18.57    &  -27:42:29.50     &  22.50   &   0.6042   &    0.604     &   6.8   &   426.9$\pm$15.0      &     184.5$\pm$4.2   \\
    123859   &   3:32:15.45    &  -27:42:20.54     &  22.68   &   0.4190   &    0.418     &   3.9   &   103.9$\pm$5.5       &      45.9$\pm$2.6   \\
    127697   &   3:32:14.74    &  -27:41:53.29     &  22.56   &   0.4170   &    0.422     &   7.0   &    21.5$\pm$4.8      &      16.1$\pm$7.1   \\
    128538   &   3:32:12.76    &  -27:41:44.45     &  22.66   &   0.4214   &    0.457     &   4.6   &    40.9$\pm$4.8      &      44.2$\pm$13.1   \\
    129968   &   3:32:11.85    &  -27:41:39.52     &  23.50   &   0.6051   &    0.603     &   3.3   &   136.3$\pm$13.9      &     190.6$\pm$22.5   \\
    130264   &   3:32:11.26    &  -27:41:27.01     &  22.30   &   1.0574   &     --       &    --   &        --            &          --          \\
    134573   &   3:32:22.01    &  -27:40:59.21     &  22.99   &   0.3579   &     --       &   8.7   &   244.1$\pm$8.2      &          --          \\
            
\footnotetext[1]{a: The optical $i$-band magnitudes are obtained from HST/ACS GOODS version 2.0 images (Giavalisco et al. 2004).}
\footnotetext[2]{b: The line FWHMs are measured for $H\beta$ and in unit of $\AA$.}
\footnotetext[3]{c: The fluxes are in unit of $10^{-18} ergs s^{-1} cm^{-2}$.}
\footnotetext[4]{d: One star marks AGN candidate identified by the CDF-S X-ray luminosity, 
hardness ratio and column density, SFRs, the BPT diagram, and the high ionization indictator emission lines.}
\footnotetext[5]{e: Two stars mark AGNs identified by the CDF-S X-ray luminosity, line widths, and spectral slope.}
\footnotetext[7]{f: The FWHM of object 92839 is measured from MgII since the H recombination lines are out of the spectral coverage.}
\footnotetext[8]{NOTE: No data indicates measurement was not possible. In case of $z_{grism}$, no data is because no suitable line ID was
found for the given input guess redshift.}

\end{longtable}
\normalsize



%% file: t2.tex



\begin{landscape}

\begin{scriptsize}
\begin{center}
\begin{table}[ht]
\begin{threeparttable}[b]
\caption[SFRs]{Star formation rates (M$_\sun$/yr) derived from line luminosities (erg/s) of [OII] and $H\beta$,
and X-ray soft-band (0.2-5 kev) and hard-band (2-10 kev) luminosities for the identified three starburst galaxies
by X-ray cross-checking. The upper limit detection is denoted.} \label{sfr} 
\begin{tabular}{cccccccccc}
\\[-2ex]
\hline \hline \\[-2ex]

PEARS ID & $z$ & $L_{[OII]}$\tablenotemark{a} & $SFR_{[OII]}$\tablenotemark{b} & $L_{H\beta}$\tablenotemark{a} & $SFR_{H\beta}$\tablenotemark{b}  & $L_{SB}$\tablenotemark{a} & $SFR_{SB}$\tablenotemark{b} & $L_{HB}$\tablenotemark{a} & $SFR_{HB}$\tablenotemark{b} 
\\[0.5ex] \hline
                                                                                                          
     59018  &    0.457   &   9.22e+40   &   1.29    &   2.92e+40   &   0.65    &   4.436e+40  &    9.76   &   $<$ 1.23e+41  &   $<$ 24.55  \\
     60143  &    0.546   &   8.54e+41   &   11.95   &         --    &     --   &   $<$ 4.411e+40  &    $<$ 9.70   &   $<$ 1.88e+41  &   $<$ 37.67   \\
     79483  &    0.435   &   1.26e+41   &   1.76    &   5.97e+40   &   1.33    &   6.632e+40  &   14.59   &   $<$ 1.31e+41  &   $<$ 26.17  \\

\hline

\end{tabular}

\begin{tablenotes}
\item[a:]{The luminosities are in unit of $ergs s^{-1}$.}
\item[b:]{The star formation rates are in unit of $M_{\sun}/yr$.}
\end{tablenotes}

\end{threeparttable}

\end{table}
\end{center}

\end{scriptsize}

\end{landscape}


%% file: ms.bbl
\begin{thebibliography}{}

\bibitem[Baldwin, Phillips, \& Terlevich(1981)]{baldwin81}Baldwin, J. A., Phillips, M. M., \& Terlevich, R. 1981, PASP, 93, 5

\bibitem[Bruzual \& Charlot (2003)]{bruzual03}Bruzual, G., \& Charlot, S. 2003, ApJ, 405, 538

\bibitem[Oemler et al.(2009, COSMOS Version 2.13)]{cosmos09}Oemler, A., Clardy, K., Kelson, D., Walth, G., Villanueva, E. 2009, COSMOS Version 2.13

\bibitem[Giavalisco et al.(2004)]{giavalisco04}Giavalisco, M., et al. 2004, ApJ, 600, 93

\bibitem[Kauffmann et al.(2003)]{kauffmann03}Kauffmann, G. et al. 2003, MNRAS, 346, 1055

\bibitem[Kennicutt(1998)]{kennicutt98}Kennicutt, J. R. 1998, ARAA, 36, 189

\bibitem[Kewley et al.(2001)]{kewley01}Kewley, L. J., Dopita, M. A., Sutherland, R. S., Heisler, C. A. \& Trevena, J. 2001, ApJ, 556, 121


\bibitem[Luo et al.(2008)]{luo08}Luo, B., et al. 2008, ApJS, 179, 19

\bibitem[Malhotra et al.(2005)]{malhotra05}Malhotra, S. 2005, HST, prop10530

\bibitem[Ranalli et al.(2003)]{ranalli03}Ranalli, P., Comastri, A., Setti, G. 2003, A\&A, 339, 39

\bibitem[Straughn et al.(2008)]{straughn08}Straughn, A. N. et al. 2008, AJ, 135, 1624

\bibitem[Straughn et al.(2009)]{straughn09}Straughn, A. N. et al. 2009, AJ, 138, 1022

\bibitem[Vanzella et al.(2006)]{vanzella06}Vanzella, E. et al. 2006, A\&A, 454, 423

\bibitem[Vanzella et al.(2008)]{vanzella08}Vanzella, E. et al. 2008, A\&A, 478, 83

\bibitem[Xu et al.(2007)]{xu07}Xu, C. et al. 2007, AJ, 134, 169

\end{thebibliography}
